# Integrating AI into Radiology workflow: Levels of research, production, and feedback maturity

Engin Dikici,[a,*] Matthew Bigelow,[a,*] Luciano M. Prevedello,[a] Richard D. White,[a] Barbaros Selnur Erdal,[a]

[a] Laboratory for Augmented Intelligence in Imaging of the Department of Radiology, The Ohio State University College of Medicine, Columbus, OH, 43210, USA

**Abstract**. This report represents a roadmap for integrating Artificial Intelligence (AI)-based image analysis algorithms into existing Radiology workflows such that: (1) radiologists can significantly benefit from enhanced automation in various imaging tasks due to AI; and (2) radiologists' feedback is utilized to further improve the AI application. This is achieved by establishing three maturity levels where: (1) *research* enables the visualization of AI-based results/annotations by radiologists without generating new patient records; (2) *production* allows the AI-based system to generate results stored in an institution's Picture Archiving and Communication System; and (3) *feedback* equips radiologists with tools for editing the AI inference results for periodic retraining of the deployed AI systems, thereby allowing the continuous organic improvement of AI-based radiology-workflow solutions. A case study (i.e., detection of brain metastases with T1-weighted contrast-enhanced 3D MRI) illustrates the deployment details of a particular AI-based application according to the aforementioned maturity levels. It is shown that the given AI application significantly improves with the feedback coming from radiologists; the number of incorrectly detected brain metastases (false positives) reduces from 14.2 to 9.12 per patient with the number of subsequently annotated datasets increasing from 93 to 217 as a result of radiologist adjudication.

**\*** Engin Dikici and Matthew Bigelow are joint first authors with equal contributions.

## 1 Introduction

Artificial Intelligence (AI) has been utilized for decades to: (1) address and solve a variety of medical imaging problems in domains such as image segmentation[1] (i.e., finding the borders of a target object), registration[2] (i.e., visually aligning anatomical parts in single- or multi- modality images), detection (i.e., detecting formations/structures), and classification[3] (i.e., grouping of medical information in subgroups); and (2) facilitate information feeds in Radiology workflows (e.g., natural language processing in dictation systems)[4,5]. Machine Learning (ML) is an application of AI where the computer ("machine") is given data access and models are used for extracting relevant information from the data. The recent usage of neural networks, which is a well-established ML approach, has gained significant momentum with generalization of the techniques to deeper network architectures referred to as Deep Neural Networks (DNNs); the complete concept is referred to as "Deep Learning"[6]. Research on deeper architectures show that the accuracy of the deployed models depends heavily on the amount of relevant information; therefore, access to both past data and the ongoing feed of new information is critical. Accordingly, Deep Learning-based solutions are commonly built on vast amount of data[7].

Medical imaging creates multiple challenges for researchers attempting to adopt Deep Learning, including the following: (1) circulation of data between institutions, or even between departments within the same institution, is complicated by various legal barriers largely related to privacy issues; (2) high-resolution and high-dimensionality (e.g., 3D+Time) of the data commonly translates into AI models with high magnitudes of parameters; large amounts of data are then needed for convergence of models; and (3) most medical-imaging applications require image annotations (i.e., segmentation and detection results) by medical experts in order to train the AI algorithms[8,9]; medical images without relevant and accessible annotations might not be useful for a variety of supervised learning scenarios in which the machine learns how to map an input image to output results by processing example input-output pairs. Thus, researchers pursuing the latest developments in ML must restructure their workflows to enable a flow of high-quality annotated information in order to both train and continuously update medical-imaging models.

Accordingly, this report introduces architectural modifications to a given Radiology workflow in multiple stages, delineating *research*, *production*, and *feedback* maturity levels. The ultimate goal of this work is to promote the integration of imaging AI into the Radiology workflow in which inference-generating models grow organically with the continuous inflow of both new medical data and radiologist feedback for ongoing model learning; hence, the *feedback* maturity level is the final goal, with *research* and *production* serving stages to achieve this end. To further solidify the understanding of these maturity levels, a case study representing the deployment of an example AI application, brain metastases detection with T1-weighted contrast-enhanced 3D MRI, is provided. The results section represents the evaluation of the accuracy of that AI application at three incremental quantities of added *feedback* data from radiologist adjudication of inference results in a simulated complete deployment (i.e., with 93, 155 and 217 annotated datasets, and using a five-fold Cross-Validation (CV)). The report concludes with a discussion of the results, system limitations, and future directions.

*1.1 Radiology Workflow and Definitions*

The implementation and maintenance of Radiology workflows have been investigated in numerous earlier studies[10,11]. The workflow example highlighted in this study is as follows (see Fig. 1):

1) Medical images are acquired with a standard modality (e.g., CT, MRI, etc.) by a technologist.

2) The images acquired in Digital Imaging and Communications in Medicine (DICOM) format, are sent to a DICOM router[12] that: (1) is capable of sending/receiving DICOM images to/from predefined addresses; and (2) contains a processing-pipeline to ensure the integrity of DICOM images that are handled.

3) The router sends the images to the: (1) Picture Archiving and Communication System (PACS); and (2) Vendor-Neutral Archive (VNA). The VNA is a technology enabling the storage of medical images in a standard format and offering a generic interface, thereby making the data accessible to multiple healthcare professionals regardless of the type of proprietary system from which images are originating[13].

4) Using dedicated workstations, radiologists access image data stored in PACS for study visualization, post-processing, and interpretation. Non-Radiology clinicians may also view medical images stored only in a VNA through links in each patient's Electronic Medical Record (EMR).

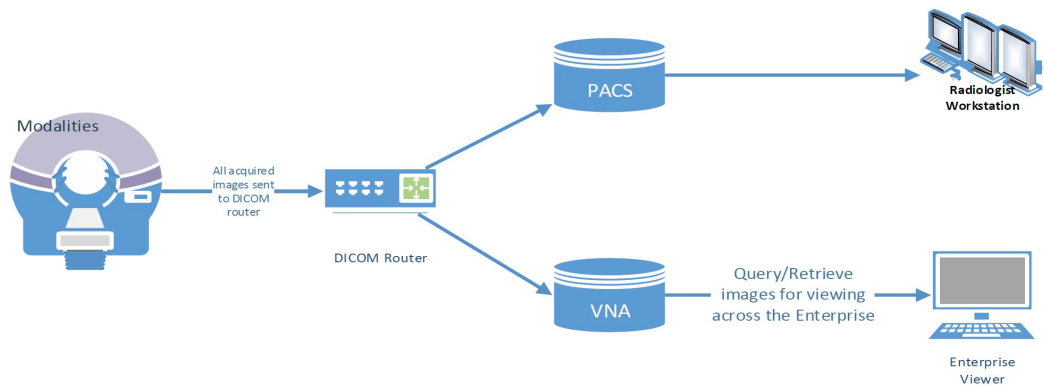

**Fig. 1** A simplified view of standard Radiology workflow: The acquired images are passed from DICOM router to the PACS and VNA. Radiologists access data stored in PACS via dedicated workstations. Other medical experts may view medical images using a VNA-directed enterprise viewer.

## 1.2 Architectural Adaptations to Achieve Levels of Maturity

### 1.2.1 Research maturity level

In order to represent the inference results from an ML algorithm to a dedicated group of medical experts (see Fig. 2), the workflow must be adapted. At this *research* maturity level, imaging modalities (e.g., CT, MRI, etc.) send acquired images to a DICOM router which then distributes received images to pertinent storage locations, such as the PACS or VNA. The images archived in PACS can then be accessed by a radiologist via dedicated workstations. Next, to benefit from the AI-based algorithm, the radiologist may send images to a DICOM node where the AI system is deployed. The AI system receives the input DICOM images, processes them, and prepares the results as a: (1) DICOM mask, where an algorithm's output is represented as another DICOM image with voxel-values representing the findings; (2) Grayscale Softcopy Presentation State (GSPS)[14]; (3) DICOM Segmentation (DICOM SEG) object[15]; or (4) DICOM Structured Report (DICOM SR) [16]. After the result dataset is generated, it is sent to a separate Research-PACS; this enables the official EMR in PACS to remain intact. The archives in the Research-PACS are accessible by standalone advanced DICOM viewers that can connect to a variety of DICOM storage locations. Advanced viewers allow radiologists and other medical experts to view and analyze AI system results related to processed DICOM images; they commonly include visualization tools to view output files (i.e., GSPS, DICOM SEG, etc.) in connection with their corresponding DICOM images.

The advanced viewers can also be configured to view images stored in a VNA. In this architecture, non-radiologists who have access to only the VNAs (with no PACS access privileges) cannot see or interact with the research AI system.

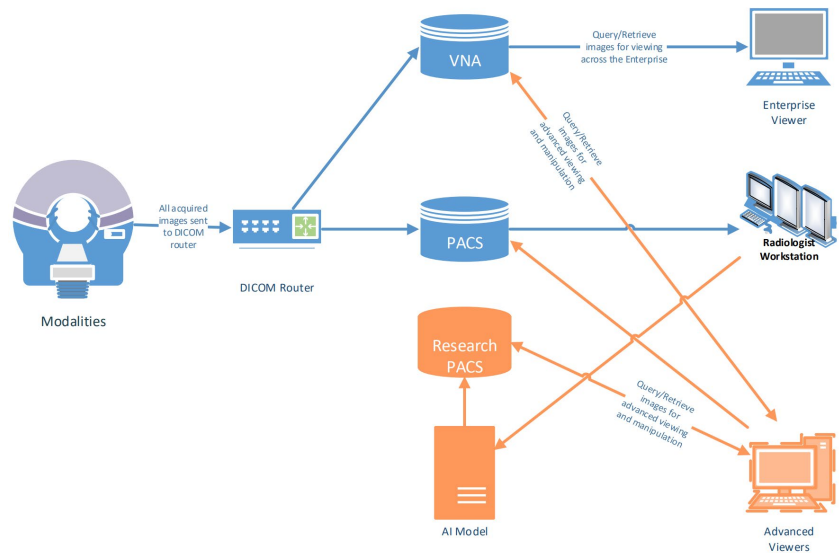

**Fig. 2** Research workflow: The medical images are acquired in various modalities, and the resulting information is passed from the DICOM router to the PACS and VNA. A radiologist can then view the data from dedicated workstations. Selected studies are passed to the AI system, and the results are viewed from advanced viewers and stored in the Research-PACS. Color coding indicates standard Radiology workflow components (blue) and additional parts required for the *research* maturity level (orange).

*1.2.2 Production maturity level*

The *production* maturity level allows verified AI models which have been deployed, optimized, and validated within the research workflow to be placed in a production mode (see Fig. 3). This introduces an additional path for acquired images to be sent directly from the DICOM router to the AI system where the received images are processed using a verified model; the results are then placed into a PACS system as a DICOM mask, GSPS, DICOM SEG, and/or DICOM SR object. Unlike at the *research* maturity level, the results become part of a patient's EMR.

The *production* maturity level enables triaging of studies based on the results from the AI model inference; it allows a study to be flagged, or accordingly prioritized, in a radiologist's worklist[17]. While this setup allows viewing of AI results in connection with their target images, receipt of radiologist *feedback* on these results is not facilitated. The *production* maturity level only aims to use the exiting AI models without allowing modifications of them.

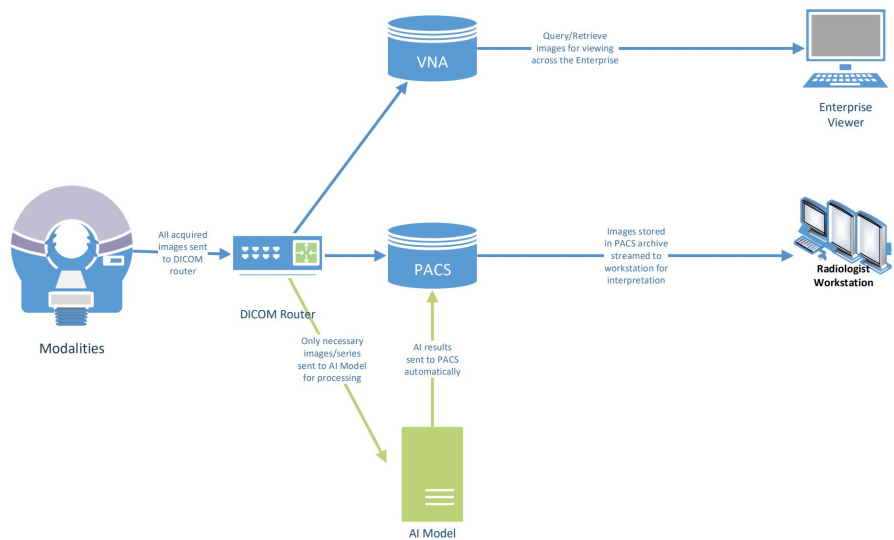

**Fig. 3** Production workflow: Once the medical images are acquired, information is passed from DICOM router to the PACS, VNA, and AI system for processing. The results of the AI system are stored in PACS as patient records, which are accessible from a radiologist's workstation. Color coding indicates standard Radiology workflow components (blue) and additional parts required for the *production* maturity level (green).

*1.2.3 Feedback maturity level*

As mentioned earlier, the accuracy of an AI model utilizing Deep Learning commonly depends on the amount of data available during the initial training. Accordingly, the *feedback* maturity level aims to place the AI model at a location where it can benefit from the constant stream of annotated data resulting from radiologist adjudication of inference results. Thus, the AI model is continuously updated/modified (see Fig. 4). This is achieved by adding a: (1) dedicated AI model training server; (2) medical-data annotation storage; and (3) medical-imaging viewer allowing adding, editing, and removal of annotations from corresponding medical images. In this workflow, a medical-imaging viewer, which can be a web-based zero-footprint version, allows: (1) visualization of the images by retrieval from PACS; (2) representation of annotations for corresponding images to be stored in a dedicated annotation storage system; and (3) editing/removal of these annotations while storing the modified annotations back to the annotation storage. The training server is set to periodically access the PACS and annotation server in order to gather newly acquired data and their annotations in order to retrain/update the AI models. The periodicity of retraining can be set depending on number of factors, such as the data acquisition frequency and workload of the training server.

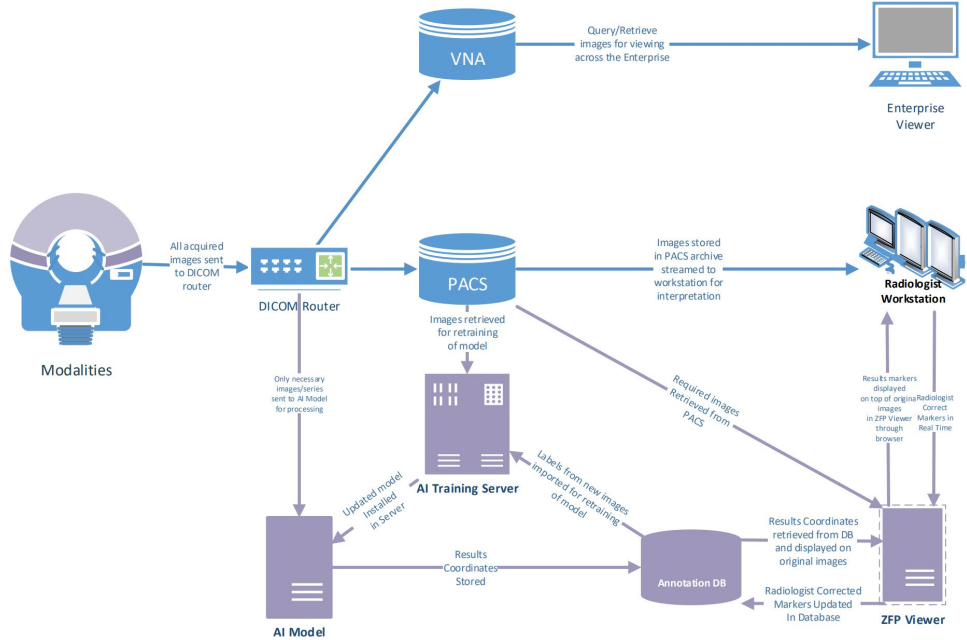

**Fig. 4** Feedback workflow: The medical images are acquired in various modalities, and the acquired information is passed from DICOM router to the PACS, VNA and AI system. A user may then access the AI results/annotations from a Zero Foot Print (ZFP) viewer and edit the results. The AI results/annotations and image data are periodically fed into a training server for updating the AI model. Color coding indicates standard Radiology workflow components (blue) and additional parts required for the *feedback* maturity level (purple).

## 2 Case Study: Brain Metastases Detection beyond CAD

Computer-Aided Detection (CAD) technology allows computational procedures to assist radiologists in the diagnosis and characterization of disease by obtaining quantitative measurements from medical images along with clinical information[18]. A classical CAD system is trained with a collection of medical-imaging datasets before deployment -as stand-alone software or as a tool in PACS and/or medical imaging viewers. The concept of integrating CADs into PACS (i.e., to allow the execution of CAD procedures on images stored in PACS) has been investigated in various previous studies[19,20].

The development of a CAD model is complete after the training procedure. However, when the model is kept up to date with future batch data updates and training(s), executed as an additional procedure which are not part of a common Radiology workflow, this additional repetitive learning opportunity for a CNN-derived model constitutes AI.

In this case study, an example AI application that detects Brain Metastases (BM) with T1-weighted contrast-enhanced 3D MRI[21] is deployed in a Radiology workflow that evolved through all three aforementioned maturity levels: *research*, *production*, and *feedback*. In the *feedback* maturity level, this AI application is beyond a common CAD; its accuracy is constantly improving due to *feedback* coming back from radiologists receiving algorithm inference results while performing clinical interpretations; *feedback* to the AI system is provided almost seamlessly using the tools that are integrated into the Radiology workflow.

*2.1 Brain Metastases Detection - Research Maturity Level*

For the given case study, the technologist acquires the T1-weighted contrast-enhanced 3D MRI data of a patient using a MRI scanner (GE Optima MR450w). The images are sent through DICOM-transfer to a DICOM router (Laurel Bridge). After the images are received by the router, they are forwarded to two different storage locations, including the institution's PACS (AGFA Impax) and VNA (Hyland). The images routed to PACS are immediately accessible by radiologists for their interpretations, whereas the images routed to the VNA are accessible to non-radiologist physicians via the patient's EMR (Epic) with the use of an enterprise viewer (Hyland NilRead).

During the medical-image interpretation, if the radiologist decides to receive inference results from the AI model, the images can be sent to the AI system via DICOM-transfer which transmits the series to the DICOM node where the AI system is located (see Fig. 5). This image transfer can be initiated at any time by the radiologist, but should ideally be done at the beginning of an interpretation in order to minimize the amount of time waiting on inference results; if images are sent as soon as the examination is opened for interpretation, the radiologist can continue viewing images while the AI model is processing a result.

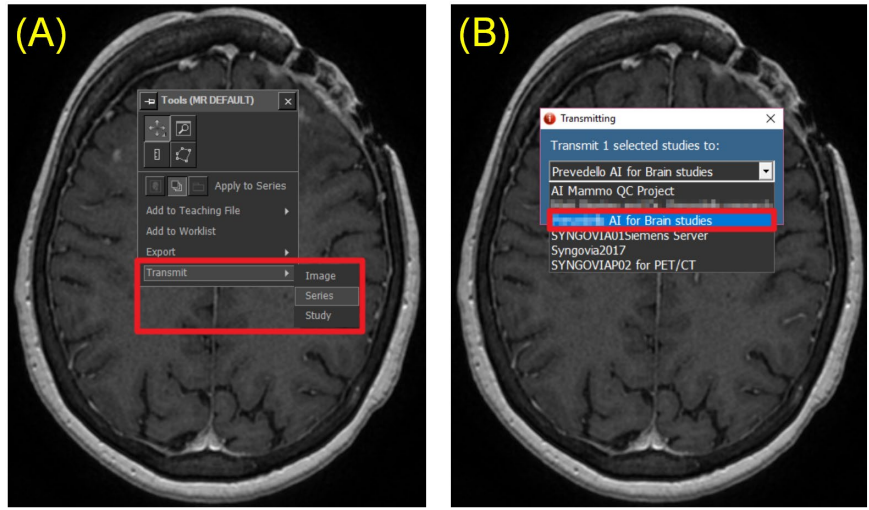

**Fig. 5** (A) DICOM-transfer can be utilized to send the MRI series, and (B) the final destination should be the DICOM node where the AI system is located.

After processing the 3D MRI data, the AI model generates a GSPS object to register its results, which are 3D coordinates of each BM-detected center position in this case. It then utilizes DICOM-transfer to send the GSPS object to the Research-PACS; in the provided setup, the results are sent to an advanced imaging analysis workstation server (Siemens SyngoVia). In the *research* architecture, this is a critical phase for keeping the results: (1) separated from the patient's EMR, and (2) inaccessible by non-radiologist personnel.

Next, the radiologist can switch to a separate viewer connected to the Research-PACS, and view the results. GSPS objects are automatically displayed over their corresponding medical images (see Fig. 6). From the Research-PACS, the radiologist switches to a different viewer, and (B) SyngoVia overlays the resulting GSPS objects on the medical images.

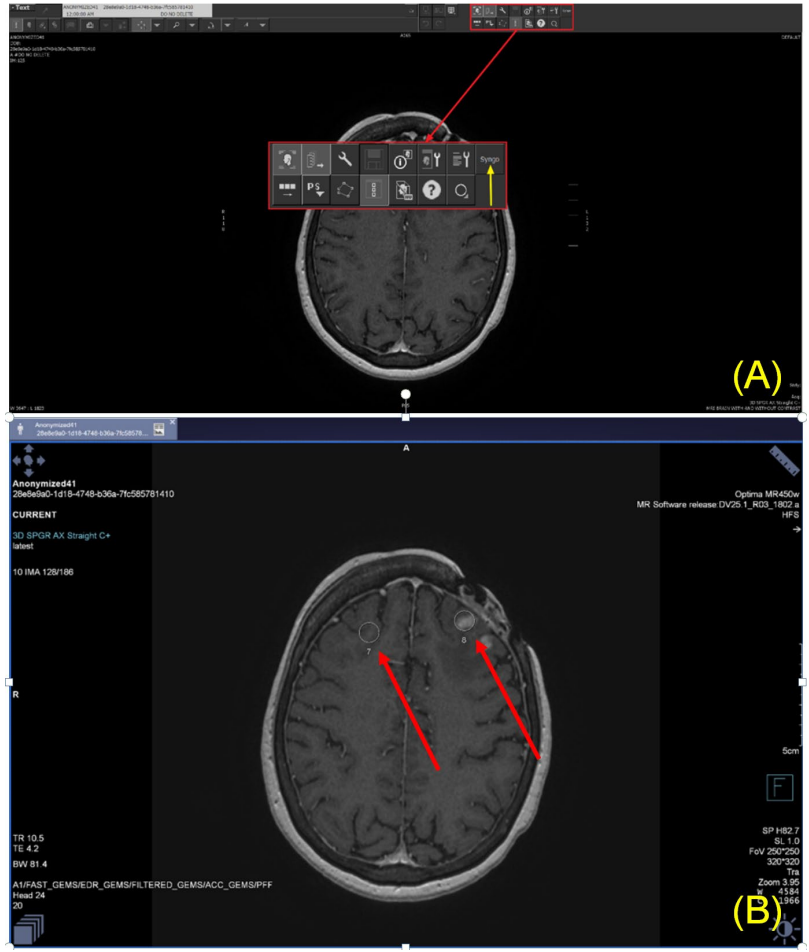

**Fig. 6** (A) From the PACS, the radiologist switches to a different viewer, and (B) SyngoVia overlays the result, GSPS object, on the medical image (GSPS circle overlays are pointed with red arrows in the figure).

## 2.2 Brain Metastases Detection - Production Maturity Level

After an AI model is approved for clinical usage, the *research* workflow can be altered slightly to achieve the *production* maturity level. First, the DICOM router is configured to concurrently send the acquired medical images directly to the institution's PACS, VNA and AI system. The routing rules of the router can be set so that it only forwards the necessary set of images to the AI system; for this case study, the pertinent images are the series for axial T1-weighted 3D MRI with contrast. For a given patient, sending a selected subset of images series, rather than a complete study, is critical; sending a complete study may take significantly more time to transmit while consuming larger storage space at the target destination.

After the AI system processes the images received from the router, it sends the results in GSPS format to the PACS server; hence, the results are available for the radiologist to view in their standard PACS workspace. The radiologist can simply load the results by selecting the AI results presentation state that displays the GSPS overlays on the corresponding MRI data (see Fig. 7).

The *production* workflow saves the radiologist time by having only the appropriate images automatically routed to the AI model, as well as the results being sent to the system where the radiologist will be utilizing them to enhance examination interpretation. The results become part of the patient's EMR in this architecture, therefore the deployed AI model must be validated properly during the *research* architecture deployment.

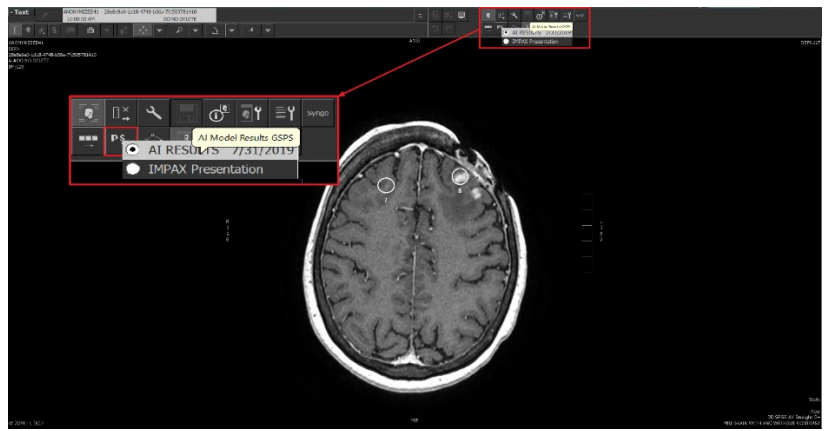

**Fig. 7** The radiologist can load the AI system's detection results from the PACS workspace by simply selecting the AI_RESULTS presentation state.

*2.3 Brain Metastases Detection - Feedback Maturity Level*

The *production* architecture can be further modified to integrate *feedback* processes into the workflow. This is achieved by first setting a viewing tool that enables the radiologist to see the AI results/annotations on their corresponding images; the viewing tool must also enable the editing of these annotations. For this purpose, a zero-footprint medical image viewer (Open Health Imaging Foundation (OHIF) Viewer)[22,23] can be modified for: (1) accessing the AI results, representing the AI detected metastases centers stored in an annotation database (MongoDB); and (2) editing/removing these results (see Fig. 8). To enable the continuous updating of the AI model, a dedicated training server (NVIDIA-DGX) can be added into the workflow, where: (1) a direct connection between PACS and the training server is established, allowing newly acquired 3D MRI datasets to be sent from PACS to the training server; and (2) the training server is also given direct access to the annotation database. By connecting the training server to both PACS and the annotation database, the training server is able to extract labeled data in any desirable format (e.g., GSPS, DICOM SEG, DICOM mask, or DICOM SR).

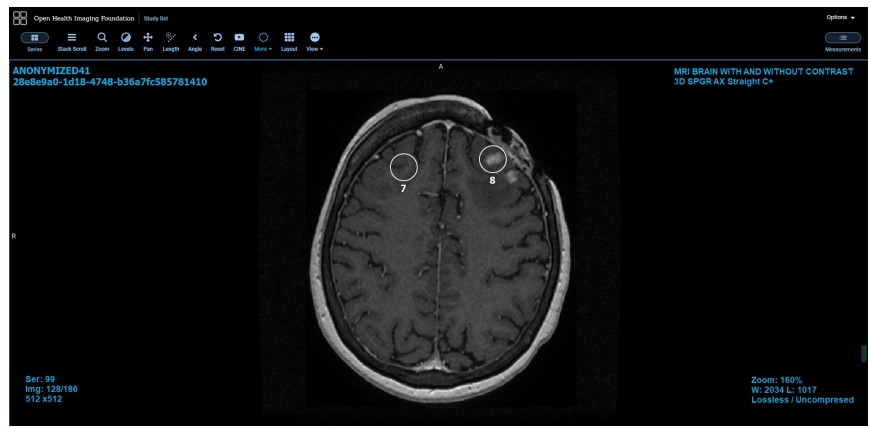

**Fig. 8** OHIF Viewer allows the radiologist to not only see the AI results but also to modify/remove them.

## 3  Results

The accuracy of the AI model used in the case study[21] is measured for the *feedback* maturity level by simulating three incremental quantities of added *feedback* data from radiologist adjudication of inference results in a simulated complete deployment. The data-selection criteria for these increments were are as follows: (1) datasets included 93 (acquired from 85 patients), 155 (from a total of 120 patients – 35 additional patients) and 217 (from a total of 158 patients – 38 additional patients) post-gadolinium T1-weighted 3D MRI exams, respectively.

The major components of this investigation, including: (1) the algorithmic details (e.g., DNN architecture, data augmentation steps, training methodology, etc.), (2) analysis of statistical properties of the BM included in the study (e.g., lesion diameter, volume, location etc.), and (3) adherence to data-acquisition criteria have been comprehensively described in a previous report [21]. This retrospective study was conducted under Institutional Review Board approval with waiver of informed consent.

The metric of Average False-Positives (AFP) per patient, representing the incorrectly detected BM lesions for each patient in relation to the sensitivity, was used during the validation of the algorithm for the three datasets[21]; the AFP values were computed using a 5-fold cross validation (see Fig. 9). At 90% sensitivity level (i.e., 90% of true BM are detected for a given test exam), the algorithm produced 14.2, 9.78, and 9.12 false positives per patient for the first (see Fig. 9 A), second (see Fig. 9 B) and third (see Fig. 9 C) datasets, respectively. The reduction of false positives from 14.2 to 9.12 with the addition of 124 exams (i.e., Dataset01 had 93, and Dataset03 had 217 exams) is a significant improvement for a BM detection system.

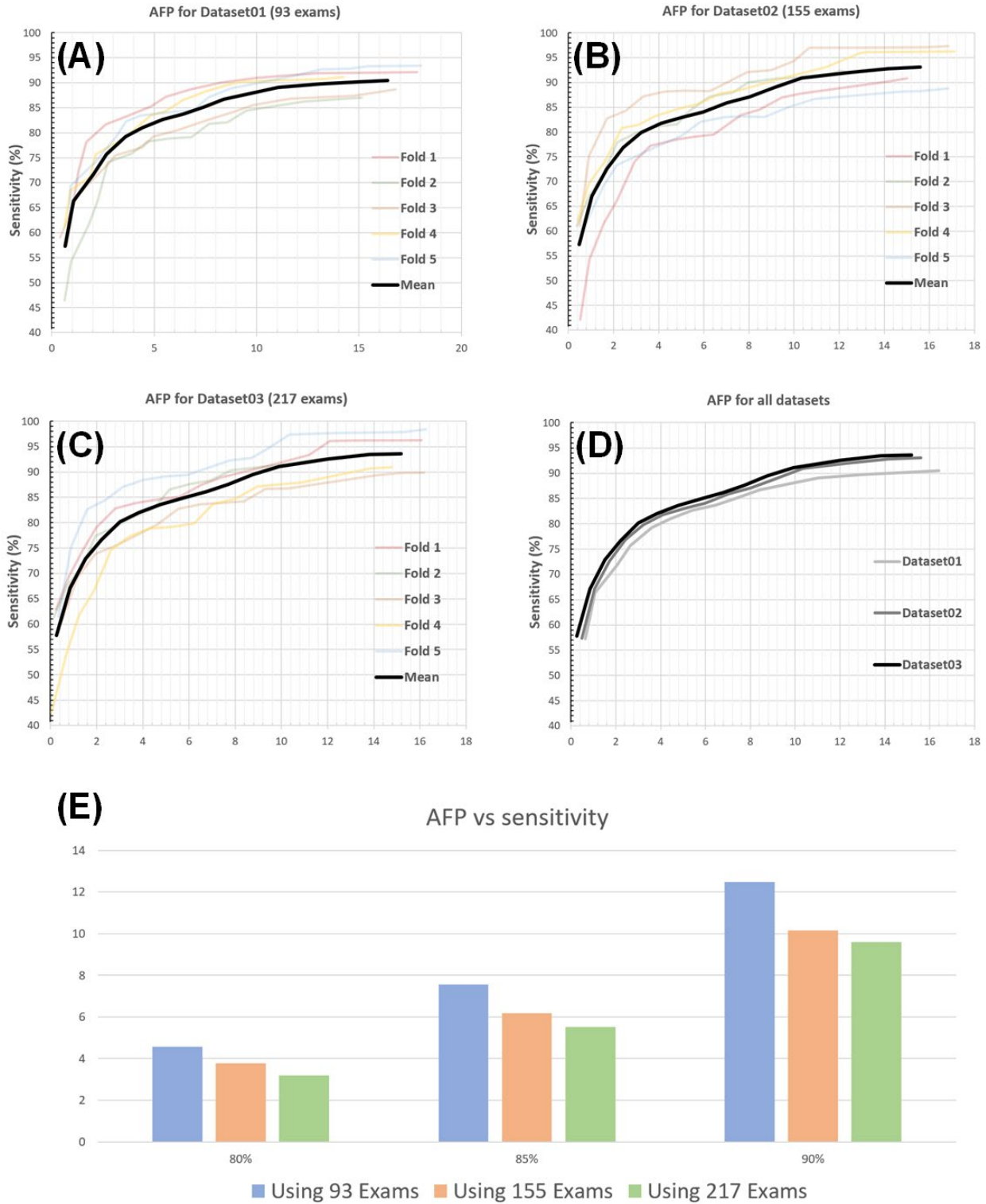

**Fig. 9** (A, B, C) Average number of false-positives per patient (i.e., wrongly detected BM lesions for each patient) in relation to the sensitivity is illustrated for each CV fold, and the mean of all folds (black curve); (D) the mean AFP curves of all datasets are shown together, with the largest resulting in the strongest performance; (E) the AFP computed for three sensitivity values (80%, 85% and 90%) for the three datasets: the AFP reduces as the number of radiologist-annotated datasets increases.

## 4 Discussion and Conclusion

It has been shown in multiple previous studies that to train more complex models it is commonly better to have more expert annotated data[24,25]. The results of this work showed that the amount of data is also a determining factor for the accuracy of the AI approach used in the given case study. However, the amount of data is not the only element that might benefit an AI-based system; the used machine learning algorithm and its parameters, as well as the properties of the added data have a significant impact on the final accuracy. While the *feedback* maturity level facilitates the gathering of annotated data and feeding of the additional data into deployed models, the selection of proper AI models is still the responsibility of the researchers/data scientists. Accordingly, it would be misleading to assume that the deployment of *feedback* architecture alone would ensure a successful AI system integration.

As ML enhances its applicability and significance in the medical-imaging domain, Radiology workflows enabling AI models to access medical data is critical. Accordingly, this report delineates three maturity levels for AI integration into a given Radiology workflow; (1) representing the results of investigational AI models to radiologists without generating new patient records; (2) processing data stored in PACS with deployed AI model, and (3) updating a deployed AI model organically via radiologist interactions with images and their annotations, which allows constant evolution of an AI model from the inference adjudication process. The case study gave implementation directions for these architectures by providing vendor options and descriptive figures.

*Disclosures*

No conflicts of interests, financial or otherwise, are declared by the authors.